\documentstyle[12pt]{article}
\begin{document}

\title{Generation of arbitrary quantum state in a high-Q cavity}
\author{XuBo Zou, K. Pahlke and W. Mathis  \\
\\Institute TET, University of Hannover,\\
Appelstr. 9A, 30167 Hannover, Germany }
\date{}

\maketitle

\begin{abstract}
{\normalsize We present a scheme to generate arbitrary
superposition of the Fock states in a high-Q cavity . This
proposal is based on a sequence of laser pulses, which are tuned
appropriately to control transitions on Fock state. It is shown
that N laser pulses are needed to generate a pure state with a
phonon number limit $N$. } PACS number(s): 03.67.-a, 03.67.Hk.
\end{abstract}
The generation of nonclassical states was extensively studied in
the past theoretically and experimentally. The first significant
advances were made in quantum optics by demonstrating antibunched
light \cite{Paul} and squeezed light \cite{Loudon}. A number of
scheme has been proposed for the purpose of generating various
special nonclassical state\cite{cb}-\cite{zou} in context of
cavity QED, trapped ion. In particular, several scheme for
generating arbitrary single mode quantum state\cite{vo}-\cite{lo}
have also been proposed and generalized to generate arbitrary two
mode quantum state\cite{sa}-\cite{zou}. Experimental realization
of highly nonclassical state, such as Fock states, squeezed and
Schr\"{o}dinger cat states of one dimensional vibrational motion
of a trapped ion and cavity field have been reported
\cite{MM,Brune}. Among of the above schemes for single mode
quantum state preparation
\begin{eqnarray} \sum_{n=0}^{N}C_n|n>\,. \label{1}
\end{eqnarray}
three basic schemes have been proposed. In the schemes proposed by
Vogel et al \cite{vo}, measurements have to be performed on the
atom in order to generate a desired pure state, which lead to
finite probability of success. The probability of detecting atom
in ground state decrease with an increase in the photon numbers.
The method described in Ref\cite{zo} is based on the adiabatic
transfer of atomic ground state Zeeman coherence, so that the
number of photon is limited by the number of available Zeeman
levels. In Ref\cite{lo}, a appropriate time-dependent cavity QED
interaction is designed to create the desired field state, which
need 2N operations to generate a pure state with a phonon number
limit $N$. The reduction of the number of operation is important
for further experimental realization. Notice that the quantum
state (\ref{1}) involves N independent complex coefficients. The
purpose of this paper is to present a scheme to generate quantum
state (\ref{1}) with N quantum operations. We will show that each
coefficient of quantum state (\ref{1}) corresponds to only one
laser pulse of this most efficient quantum state generation
scheme.\\
In order to describe our scheme, we consider the a high-Q cavity
partially filled by a Kerr medium with a series of coherent pulses
injected into it, whose Hamiltonian is given by\cite{jr}
$$
H=\Delta_{\omega}a^{\dagger}a+\chi a^{\dagger2}a^2+
ge^{i\varphi}a^{\dagger}+ge^{-i\varphi}a \eqno{(2)}
$$
Here $a^{\dagger}a$ and $a$ are creative and annihilation
operators. The parameters $\Delta_{\omega}$ and $\chi$ describe
the detuning of the cavity from the driving frequency and the Kerr
nonlinearity strength, respectively. The $\varphi$ and g are phase
and amplitude of driving laser field, respectively. From a
practical point of view, parameter $\Delta_{\omega}$, $\varphi$
and g are adjustable in the experiment to give control over the
cavity field. It is noticed that Kerr nonlinear has been used to
generate Fock state from vacuum state in Ref\cite{jr}. Recently an
optical Fock-state synthesizer was proposed to synthesize Fock
state and their superpositions from a coherent state\cite{go}.
Here we present a scheme to generate arbitrary quantum state(1)
via adjusting frequency of driving laser field. We choose
frequency of laser driving field to make detuning
$\Delta_{\omega}$ to satisfy the condition
$\Delta_{\omega}=2i\chi$, $i=0,1,\cdots,N-1$. Thus Hamiltonian of
the system can be written in the form
$$
H=\chi(a^{\dagger}a-i)(a^{\dagger}a-i-1)+ge^{i\varphi}a^{\dagger}+ge^{-i\varphi}a
\eqno{(3)}
$$
Here we have neglected a constant term.  Using the Fock state
$|n>$, $n=0,1\cdots$ the Hamiltonian(3) take the form
$$
H=\sum_{n}\chi(n-i)(n-i-1)|n><n|
+g_ie^{i\varphi_i}\sum_{n=0}\sqrt{n+1}|n+1><n|
$$
$$
+g_ie^{-i\varphi_i}\sum_{n=0}\sqrt{n+1}|n><n+1| \eqno{(4)}
$$
It is noticed that the coefficients of the Fock state $|i>$ and
$|i+1>$ are equal to zero in the first line of above equation. We
now consider the condition $\chi>>Jg$. In this case, we apply the
rotating wave approximation and discard the rapidly oscillating
term in the Hamiltonian (4) and obtain the effective interaction
$$
H=e^{i\varphi_i}\Omega_i|i+1><i| +e^{-i\varphi_i}\Omega_i|i><i+1|
\eqno{(5)}
$$
here $\Omega_i=g_i\sqrt{i+1}$. In order to generate any
superposition of Fock state(1), we consider the situation in which
the cavity field is prepared in the vacuum state
$$
\Psi_{initial}=|0> \eqno{(6)}
$$
In the following we show, that each term of equation(1) can be
generated by one laser pulse. Fist, we derive the cavity field
with laser frequency $\Delta_{\omega}=0$, after an interaction
time $\tau_0$, the system evolves into
$$
\Psi_{1}=\cos(\Omega_0\tau_0)|0>+i\sin(\Omega_0\tau_0)e^{\varphi_0}|1>
\eqno{(7)}
$$
we choose the amplitude (or interaction time $\tau_0$) of the
laser field in such a way that the following conditional are
fulfilled
$$
\cos(\Omega_0\tau_0)=C_{0} \eqno{(8)}
$$
we obtain
$$
\Psi_{1}=C_{0}|0>+ie^{\varphi_0}\sqrt{1-C_{0}^2}|1> \eqno{(9)}
$$
without loss of generality, we have assumed that $C_{0}$ is a real
number. We then tune the laser frequency to satisfy
$\Delta_{\omega}=2\chi$. After an interaction time $\tau_1$, the
system is evolved into
$$
\Psi_{2}=C_{0}|0>+ie^{\varphi_0}\sqrt{1-C_{0}^2}(\cos(\Omega_1\tau_1)|1>+i\sin(\Omega_1\tau_1)e^{\varphi_1}|2>)
\eqno{(10)}
$$
we adjust the amplitude (or interaction time $\tau_1$) of the
second laser field and phase of the laser field of the first laser
field to satisfy
$$
ie^{\varphi_0}\sqrt{1-C_{0}^2}\cos(\Omega_1\tau_1)=C_{1}
\eqno{(11)}
$$
the state become
$$
\Psi_{2}=C_{0}|0>+C_{1}|1>+i^2e^{i\varphi_0+i\varphi_1}\sqrt{1-C_{0}^2-|C_{1}|^2}|2>
\eqno{(12)}
$$
If this procedure is done for the $m-$th time, the quantum state
of the system is
$$
\Psi_{m}=\sum_{l=0}^{m-1}C_l|l>+i^m\exp(i\sum_{l=0}^{m-1}\varphi_l)\sqrt{1-\sum_{l=0}^{m-1}|C_1|^2}|m>
\eqno{(13)}
$$
We now consider the $(m+1)-$th operation by choosing
$\Delta_{\omega}=2m\chi$. After interaction $\tau_m$, the quantum
state becomes
$$
\Psi_{m+1}=\sum_{l=0}^{m-1}C_l|J,l>+i^m\exp(i\sum_{l=0}^{m-1}\varphi_l)\sqrt{1-\sum_{l=0}^{m-1}|C_1|^2}(
\cos(\Omega_m\tau_m)|m>
$$
$$
+i\sin(\Omega_m\tau_m)e^{\varphi_m}|m+1>) \eqno{(14)}
$$
We choose the amplitude (or interaction time $t_{m}$) of the
(m+1)-th laser pulse and phase of m-th laser field to satisfy
$$
i^m\exp(i\sum_{l=0}^{m-1}\varphi_l)\sqrt{1-\sum_{l=0}^{m-1}|C_1|^2}
\cos(\Omega_m\tau_m)=C_{m} \eqno{(15)}
$$
After the procedure is performed for $N$ times the system's state
definitely becomes state(1).\\
One of the difficulties of our scheme in respect to an
experimental demonstration consists in the requirement on the
large Kerr coupling coefficient. Recently,two approaches have been
suggested to strongly enhance nonlinearlity- cavity quantum
electrodynamics\cite{ao} and electromagnetically induced
transparency\cite{a1}. These result indicate that giant Kerr
nonlinearlity can beobtained by methods not too far from present
technology. Recently a experiment have been observed that decay
time of cavity field is 0.2s\cite{a2}, which is a long time enough
time to to store field state. We present a scheme to generate
arbitrary superposition of the Fock states of the electromagnetic
field in high-Q cavity partially filled by a Kerr medium with a
series of coherent pulses injected into it. This proposal is based
on a sequence of laser pulses, which are tuned appropriately to
control transitions on Fock state. It is shown that N laser pulses
are needed to generate a pure state with a phonon number limit
$N$.


\begin{thebibliography}{99}
\bibitem{Paul} See, for a review, H. Paul, Rev. Mod. Phys. {\bf
54}, 1061 (1982); R. Loudon, Rep. Prog. Phys. {\bf43}, 913 (1980).
\bibitem{Loudon} See, for a review, R. Loudon and P. L. Knight, J.
Mod. Opt. {\bf 34}, 709 (1987).
\bibitem{cb}J. I. Cirac, R. Blatt, A. S. Parkins, and P. Zoller, Phys. Rev.
Lett. 70, 762 (1993).
\bibitem{cp}J. I. Cirac, A. S. Parkins, R. Blatt, and P. Zoller, Phys. Rev. Lett. 70, 556 (1993).
\bibitem{de}R. L. de Matos Filho and W. Vogel, Phys. Rev. Lett. 76, 608 (1996).
\bibitem{jf}J. F. Poyatos, J. I. Cirac, R. Blatt, and P. Zoller, Phys. Rev. A 54, 1532 (1996).
\bibitem{jr}J. R. Kuklinski, Phys. Rev. Lett. 64, 2507 (1990)
\bibitem{g0}P. A. Khandokhin, E. A. Ovchinnikov, and E. Yu. Shirokov
,Phys. Rev. A 61, 053807 (2000)
\bibitem{sc}S. C. Gou, J. Steinbach, and P. L. Knight, Phys. Rev. A 54, R1014 (1996).
\bibitem{cc}C. C. Gerry, S. C. Gou, and J. Steinbach, Phys. Rev. A 55, 630 (1997).
\bibitem{xb}XuBo Zou, Jaewan Kim, Hai-Woong Lee, Phys. Rev. A 63, 065801 (2001)
\bibitem{vo}K. Vogel, V. M. Akulin, and W. P. Schleich, Phys. Rev. Lett. 71,
1816 (1993).
\bibitem{zo}A. S. Parkins, P. Marte, P. Zoller, and H. J. Kimble, Phys. Rev. Lett. 71, 3095 (1993)
\bibitem{lo}C. K. Law and J. H. Eberly, Phys. Rev. Lett. 76, 1055–1058 (1996)
\bibitem{sa}S. A. Gardiner, J. I. Cirac, and P. Zoller, Phys. Rev. A 55,
1683 (1997).
\bibitem{kc}B. Kneer and C. K. Law, Phys. Rev. A 57, 2096 (1998).
\bibitem{dh}G. Drobny, B. Hladky, and V. Buzek, Phys. Rev. A 58, 2481 (1998)
\bibitem{ZH} S. B. Zheng, Phys. Rev. A {\bf 63}, 015801 (2000).
\bibitem{zou}XuBo Zou, K. Pahlke and W. Mathis, to appear in Phys.
Rev. A.
\bibitem{Brune} M. Brune, E. Hagley, J. Dreyer, X. Maitre, A.
Maali, C. Wunderlich, J. M. Raimond, and S. Haroche, Phys. Rev.
Lett. {\bf 77}, 4887 (1996).
\bibitem{MM} D. M. Meekhof, C. Monroe, B. E. King, W. M. Itano,
and D. J. Wineland, Phys. Rev. Lett. {\bf 76}, 1796 (1996); C.
Monroe, D. M. Meekhof, B. E. King, and D. J. Wineland, Science
{\bf 272}, 1131 (1996).
\bibitem{ao}
Q.A. Turchette, C.J. Hood, W. Lange, H. Mabuchi, and H.J. Kimble,
Phys. Rev. Lett. 75, 4710 (1995). J.-F. Roch, K. Vigneron, Ph.
Grelu, A. Sinatra, J.-Ph. Poizat, and Ph. Grangier, Phys. Rev.
Lett. 78, 634 (1997). C.J. Hood, M.S. Chapman, T.W. Lynn, and H.J.
Kimble, Phys. Rev. Lett. 80, 4157 (1998).
\bibitem{a1}
S.E. Harris, J.E. Field, and A. Imamolu, Phys. Rev. Lett. 64, 1107
(1990). H. Schmidt and A. Imamolu, Opt. Lett. 21, 1936 (1996).
M.D. Lukin and A. Imamolu, Phys. Rev. Lett. 84, 1419 (2000).
\bibitem{a2}B. T. H. Varcoe, S. Brattke, M. Weidinger, H. Walther,
Nature403, 743(2000)

\end{thebibliography}
\end{document}